\documentstyle[12pt,fleqn]{article}

\font\twlgot =eufm10 scaled \magstep1
\font\egtgot =eufm8
\font\sevgot =eufm7
\font\twlmsb =msbm10 scaled \magstep1
\font\egtmsb =msbm8
\font\sevmsb =msbm7

\newfam\gotfam

\textfont\gotfam\twlgot
\scriptfont\gotfam\egtgot
\scriptscriptfont\gotfam\sevgot

\newfam\msbfam
\textfont\msbfam\twlmsb
\scriptfont\msbfam\egtmsb
\scriptscriptfont\msbfam\sevmsb
\def\Bbb{\protect\pBbb}
\def\pBbb{\relax\ifmmode\expandafter\Bb\else\typeout{You cann't use
Bbb in text mode}\fi}
\def\Bb #1{{\fam\msbfam\relax#1}}

\def\thebibliography#1{\section*{References}\list
   {[\arabic{enumi}]}{\settowidth\labelwidth{#1}\leftmargin\labelwidth
     \advance\leftmargin\labelsep
     \usecounter{enumi}}
     \def\newblock{\hskip .11em plus .33em minus .07em}
     \sloppy\clubpenalty4000\widowpenalty4000
     \sfcode`\.=1000\relax}

\def\op#1{\mathop{\fam0 #1}\limits}

\newcommand{\beq}{\begin{equation}}
\newcommand{\eeq}{\end{equation}}
\newcommand{\ben}{\begin{eqnarray}}
\newcommand{\een}{\end{eqnarray}}
\newcommand{\be}{\begin{eqnarray*}}
\newcommand{\ee}{\end{eqnarray*}}
\newcommand{\bea}{\begin{eqalph}}
\newcommand{\eea}{\end{eqalph}}

\newcommand{\cF}{{\cal F}}

\newcommand{\bL}{{\bf L}}

\newcommand{\dl}{\delta}
\newcommand{\la}{\lambda}

\newcommand{\g}{\gamma}

\newcommand{\ol}{\overline}
\newcommand{\dr}{\partial}

\newcommand{\ve}{\varepsilon}

\newcounter{eqalph}
\newcounter{equationa}
\newcounter{theorem}
\newcounter{remark}
\newcounter{proposition}
\newcounter{lemma}
\newcounter{corollary}
\newcounter{definition}

\newenvironment{eqalph}{\stepcounter{equation}
\setcounter{equationa}{\value{equation}}
\setcounter{equation}{0}

\begin{eqnarray}}{\end{eqnarray}\setcounter{equation}{\value{equationa}}}

\def\theremark{\arabic{remark}}

\def\thedefinition{\arabic{definition}}

\newenvironment{proof}{\noindent
{\bf Proof.}}{\hfill $\Box$ \medskip}

\newenvironment{prop}{\refstepcounter{definition}
\medskip\noindent{\bf Proposition \thedefinition.}\it }{\medskip}

\newenvironment{cor}{\refstepcounter{definition}
\medskip\noindent{\bf Corollary \thedefinition.}\it }{\medskip}

\newcommand{\mar}[1]{}

\begin{document}
\hbox{}

{\parindent=0pt

{\large\bf The Lyapunov stability of first order dynamic equations
with respect to time-dependent Riemannian metrics}
\medskip

{\sc Gennadi 
Sardanashvily\footnote{E-mail address: sard@grav.phys.msu.su}}
\medskip

\begin{small}
Department of Theoretical Physics, Physics Faculty, Moscow State
University, 117234 Moscow, Russia
\medskip

{\bf Abstract.}
Solutions of a
smooth first order dynamic equation can be made
Lyapunov stable at will by the choice of an
appropriate time-dependent Riemannian metric.
\end{small}
}

\section{Introduction}

The notion of the Lyapunov stability of a dynamic equation on a smooth manifold
implies that this manifold is equipped with a Riemannian metric.
At the same time, no preferable Riemannian metric is associated to a
first order dynamic
equation.  
Here, we aim to study the Lyapunov stability of first order
dynamic equations in non-autonomous mechanics
with respect to different (time-dependent) Riemannian metrics. 

Let us recall that a solution $s(t)$, $t\in\Bbb R$, of a first order
dynamic equation 
is said to be Lyapunov stable (in the positive direction) if for 
$t_0\in\Bbb R$ and any
$\ve>0$, there is $\dl>0$ such that, if $s'(t)$ is
another solution and the distance between the points $s(t_0)$ and $s'(t_0)$ is
inferior to $\dl$, then the distance between the points $s(t)$ and $s'(t)$
for all $t>t_0$ is inferior to $\ve$. In order to formulate a criterion
of the Lyapunov stability with respect to a time-dependent Riemannian
metric, we introduce the notion of a covariant Lyapunov tensor as
generalization of the well-known Lyapunov matrix.
The latter is defined as the coefficient matrix of
the variation equation \cite{gal,hirs}, and fails to be a tensor under
coordinate transformations, unless they are linear
and time-independent. On the contrary, the covariant Lyapunov tensor is
a true tensor 
field, but it  
essentially depends on the choice of a Riemannian metric.
We show the following (see Propositions \ref{ch40}, \ref{ch75}, and
\ref{ch90} below).

(i) If the covariant Lyapunov tensor is negative definite in a tubular
neighbourhood of a solution $s$ at points $t\geq t_0$, this solution is
Lyapunov stable.

(ii) For any first order dynamic equation, there exists a (time-dependent)
Riemannian  metric such that every solution of this equation is
Lyapunov stable.

(iii) Moreover, the Lyapunov exponent of any solution of a first order
dynamic equation can be made equal to any real number
with respect to the appropriate
(time-dependent) Riemannian metric. It follows that
chaos in dynamical systems described by smooth ($C^\infty$) first
order dynamic equations can be characterized in full by time-dependent
Riemannian metrics.

\section{Geometry of first order dynamic equations}

Let $\Bbb R$ be the time axis provided with the Cartesian coordinate $t$
and transition functions $t'=t+$const.
In geometric terms \cite{book98}, a (smooth) first order dynamic
equation in non-autonomous 
mechanics is defined as
a vector field $\g$ on a smooth fibre bundle
\mar{ch0}\beq
\pi:Y\to\Bbb R \label{ch0}
\eeq
which obeys the condition
$\g\rfloor dt=1$, i.e.,
\mar{ch3}\beq
\g=\dr_t + \g^k\dr_k. \label{ch3}
\eeq
The associated first order dynamic
equation takes the form
\mar{ch5}\beq
\dot t=1, \qquad \dot y^k=\g^k(t,y^j)\dr_k, \label{ch5}
\eeq
where $(t,y^k,\dot t,\dot y^k)$ are holonomic coordinates on $TY$.
Its solutions are trajectories of the vector field
$\g$ (\ref{ch3}). They assemble into
a (regular) foliation $\cF$ of $Y$. Equivalently, $\g$ (\ref{ch3})
is defined as a connection on the fibre bundle (\ref{ch0}).

A fibre bundle $Y$ (\ref{ch0}) is trivial, but it admits different
trivializations
\mar{ch4}\beq
Y\cong\Bbb R\times M, \label{ch4}
\eeq
distinguished by fibrations $Y\to M$. 
For instance, if there is a trivialization (\ref{ch4})
such that,
with respect to the associated coordinates,
the components
$\g^k$ of the connection $\g$ (\ref{ch3}) are independent of $t$,
one says that $\g$ is a conservative first order dynamic equation
on $M$.

Hereafter, the vector field $\g$ (\ref{ch3}) is assumed to be complete,
i.e., there is a unique global
solution of the dynamic equation $\g$ through each point of $Y$.
For instance, if fibres of $Y\to\Bbb R$ are compact, any vector field 
$\g$ (\ref{ch3}) on $Y$ is complete.

\begin{prop} \label{ch8} \mar{ch8}
If the vector field $\g$ (\ref{ch3}) is complete,
there exists a trivialization
(\ref{ch4}) of $Y$ such that any solution $s$ of $\g$ reads
\be
s^a(t)={\rm const}., \qquad t\in \Bbb R,
\ee
with respect to associated bundle coordinates $(t,y^a)$.
\end{prop}

\begin{proof}
If $\g$ is complete, the foliation $\cF$ of its trajectories
is a fibration $\zeta$ of $Y$ along these trajectories onto any fibre
of $Y$, e.g., $Y_{t=0}\cong M$. This fibration yields a desired trivialization
\cite{book98}.
\end{proof}

One can think of the coordinates $(t,y^a)$ in Proposition \ref{ch8} as being
the initial date coordinates because all points of the same
trajectory differ from each other only in the temporal coordinate.

Let us consider the canonical lift $V\g$ of the vector
field $\g$ (\ref{ch3}) onto the vertical tangent bundle $VY$ of $Y\to\Bbb R$.
With respect to the holonomic bundle coordinates $(t,y^k,\ol y^k)$
on $VY$, it reads
\be
V\g= \g  +
\dr_j\g^k \ol y^j\ol\dr_k, \qquad \ol\dr_k=\frac{\dr}{\dr \ol y^k}.
\ee
This vector field obeys the condition $V\g\rfloor dt=1$, and
defines the first order dynamic equation
\mar{ch12}\bea
&& \dot t=1, \qquad \dot y^k=\g^k(t,y^i), \label{ch12a} \\
&& \dot {\ol y_t}^k=
\dr_j\g^k(t,y^i)\ol y^j\label{ch12b}
\eea
on $VY$. The equation (\ref{ch12a}) coincides with the initial one (\ref{ch5}).
The equation (\ref{ch12b}) is the well-known variation
equation.
Substituting a solution $s$ of the initial dynamic equation
(\ref{ch12a}) into (\ref{ch12b}), one obtains a linear dynamic equation
whose  solutions $\ol s$ are Jacobi fields of
the solution $s$. In particular, if $Y\to\Bbb R$ is
a vector bundle, there are the canonical splitting $VY\cong Y\times Y$ and
the morphism $VY\to Y$ so that
$s +\ol s$ obeys the initial dynamic
equation (\ref{ch12a}) modulo the terms of order $>1$ in $\ol s$.

\section{The covariant Lyapunov tensor}

The collection of coefficients
\mar{ch70}\beq
l_j{}^k=\dr_j\g^k \label{ch70}
\eeq
of the variation equation (\ref{ch12b}) is called the Lyapunov matrix.
Clearly, it is not a tensor under
bundle coordinate transformations of the fibre bundle $Y$ (\ref{ch0}).
Therefore, we introduce a covariant Lyapunov tensor as follows.

Let a fibre bundle $Y\to\Bbb R$ be provided with a Riemannian fibre metric
$g$, defined as a section of the symmetrized tensor product
\mar{hh1}\beq
\op\vee^2 V^*Y\to Y \label{hh1}
\eeq
of the
vertical cotangent bundle $V^*Y$ of $Y\to\Bbb R$. With respect to the
holonomic coordinates $(t,y^k,\ol y_k)$ on $V^*Y$,
it takes the coordinate form
\be
g=\frac12 g_{ij}(t,y^k)\ol dy^i\vee \ol dy^j, 
\ee
where $\{\ol dy^i\}$ are the holonomic fibre bases for $V^*Y$.

Given a first order dynamic equation $\g$, let
\mar{ch13}\beq
V^*\g= \g -
\dr_j\g^k \ol y_k\ol\dr^j, \qquad \ol\dr^j=\frac{\dr}{\dr \ol y_j}.
\label{ch13}
\eeq
be the canonical lift of the vector field $\g$ (\ref{ch3}) onto $V^*Y$.
It is a connection on $V^*Y\to\Bbb R$. Let us consider the Lie derivative
$\bL_\g g$ of the Riemannian fibre metric $g$ along the vector
field $V^*\g$ (\ref{ch13}). It reads
\mar{ch14}\beq
L_{ij}=(\bL_\g g)_{ij}= \dr_t g_{ij}+ \g^k\dr_k g_{ij}
+ \dr_i\g^k g_{kj} + \dr_j\g^k g_{ik}. \label{ch14}
\eeq
This is a section of the fibre bundle (\ref{hh1}) and, consequently, a
tensor with respect
to any bundle coordinate
transformation of the fibre bundle (\ref{ch0}).
We agree to call it the covariant Lyapunov tensor.
If $g$ is an Euclidean metric, it comes to symmetrization
\be
  L_{ij}=\dr_i\g^j + \dr_j\g^i=l_i{}^j +l_j{}^i
\ee
of the Lyapunov matrix (\ref{ch70}).

Let us point the following two properties of the covariant
Lyapunov tensor.

(i) Written with respect to the initial date coordinates in
Proposition \ref{ch8},
the covariant Lyapunov tensor is
\be
L_{ab}= \dr_t g_{ab}. 
\ee

(ii) Given a solution $s$
of the dynamic equation $\g$ and
a solution $\ol s$ of the variation equation (\ref{ch12b}), we have
\be
L_{ij}(t,s^k(t))\ol s^i\ol s^j=
\frac{d}{dt}(g_{ij}(t,s^k(t))\ol s^i\ol s^j).
\label{ch71}
\ee

The definition of the covariant Lyapunov tensor (\ref{ch14})
depends on the choice of a Riemannian fibre metric on the
fibre bundle $Y$.

\begin{prop} \label{ch73} \mar{ch73}
If the vector field $\g$ is complete, there is a Riemannian fibre metric
on $Y$ such that the covariant Lyapunov tensor vanishes everywhere.
\end{prop}

\begin{proof}
Let us choose the atlas of the initial date coordinates in
Proposition \ref{ch8}. Using the fibration $\zeta:Y\to Y_{t=0}$,
one can provide $Y$ with a time-independent
Riemannian fibre metric
\mar{ch80}\beq
g_{ab}(t,y^c)=h(t)g_{ab}(0,y^c) \label{ch80}
\eeq
  where
$g_{ab}(0,y^c)$ is a Riemannian metric on the fibre $Y_{t=0}$ and $h(t)$
is a positive smooth function on $\Bbb R$. The covariant
Lyapunov tensor with respect to the metric (\ref{ch80})
is
\be
L_{ab}=\dr_t h g_{ab}.
\ee
Putting $h(t)=1$, we obtain $L=0$.
\end{proof}

\section{The Lyapunov stability of a first order dynamic equation}

With the covariant Lyapunov tensor (\ref{ch14}), we obtain the
following criterion of the stability condition of Lyapunov.

Recall that, given a Riemannian fibre metric $g$ on a fibre bundle
$Y\to\Bbb R$, 
the instantwise distance $\rho_t(s,s')$
between two solutions $s$ and $s'$
of a dynamic equation $\g$ on $Y$ at an instant $t$
is the distance between the points $s(t)$ and $s'(t)$
in the Riemannian space $(Y_t,g(t))$.

\begin{prop} \label{ch40} \mar{ch40}
Let $s$ be a solution
of a first order dynamic equation $\g$. If there exists
an open tubular neighbourhood $U$ of the trajectory $s$
where the covariant Lyapunov tensor
(\ref{ch14}) is negative-definite at all instants $t\geq t_0$, then
there exists an open tubular neighbourhood $U'$ of $s$ such that
\be
\lim_{t'\to\infty}[\rho_{t'}(s,s')-\rho_t(s,s')]<0
\ee
for any $t>t_0$ and any solution $s'$ crossing $U'$.
\end{prop}

\begin{proof}
Since the condition and the statement of Proposition 
are coordinate-independent, let us choose the following chart of
initial date coordinates in Proposition \ref{ch8} which
cover the trajectory $s$. 
Put $t=0$ without a loss of generality. There is an open 
neighbourhood $U_0\subset
Y_0\cap U$ of $s(0)$ in the Riemannian manifold $(Y_0,g(0))$ 
which can be provided
with the normal coordinates $(x^a)$ defined by
the Riemannian metric
$g(0)$ in $Y_0$ and centralized at $s(0)$. Let us
consider the open tubular $U'=\zeta^{-1}(U_0)$ endowed with the
coordinates $(t,x^a)$. It is the desired chart of
initial date coordinates. 
With respect to these coordinates, the solution
$s$ reads $s^a(t)=0$. Let
\be
s'^a(t)=u^a={\rm const}. 
\ee
be another solution crossing $U'$.
The instantwise distance $\rho_t(s,s')$, $t\geq 0$, between solutions $s$ and
$s'$ is the distance between the points
$(t,0)$ and $(t,u)$ in the Riemannian space $(Y_t,g(t))$.
This distance does not
exceed the length
\mar{ch31}\beq
\ol\rho_t(s,s')=\left[\op\int^1_0 g_{ab}(t,\tau u^c)
u^a u^b d\tau\right]^{1/2}
\label{ch31}
\eeq
of the curve
\mar{ch34}\beq
x^a=\tau u^a, \qquad \tau\in [0,1] \label{ch34}
\eeq
in the Riemannian space $(Y_t,g(t))$, while
\be
\rho_0(s,s')=\ol \rho_0(s,s')
\ee
The temporal derivative of the function
$\ol\rho_t(s,s')$ (\ref{ch31}) reads
\mar{ch32}\beq
\dr_t\ol\rho_t(s,s')= \frac{1}{2(\ol\rho_t(s,s'))^{1/2}}
\op\int^1_0 \dr_tg_{ab}(t,\tau u^c)
u^a u^b d\tau. \label{ch32}
\eeq
Since the bilinear form $\dr_tg_{ab}=L_{ab}$, $t\geq 0$, is negative-definite
at all points of the curve (\ref{ch34}), the derivative
(\ref{ch32}) at all points $t\geq t_0$ is also negative. Hence,
we obtain
\be
\rho_{t>0}(s,s')<\ol \rho_{t>0}(s,s')<\ol \rho_0(s,s')= \rho_0(s,s').
\ee
\end{proof}

\begin{cor} \label{ch41} \mar{ch41}
The solution $s$ obeying the condition of Proposition \ref{ch40} is
Lyapunov stable with respect to the Riemannian fibre metric $g$.
\end{cor}

It is readily observed that Proposition \ref{ch40} states something more.
One can think of the solution $s$ in Proposition \ref{ch40} as being 
isometrically Lyapunov stable. Of course, being Lyapunov stable with
respect a Riemannian fibre metric $g$, a solution $s$ need not be so
with respect to another Riemannian fibre metric $g'$, unless $g'$
results from $g$ by a time-independent transformation.

\begin{prop} \label{ch75} \mar{ch75}
For any first order dynamic equation defined by
a complete vector field $\g$ (\ref{ch3}) 
on a fibre bundle $Y\to \Bbb R$,
there exists a Riemannian fibre metric on $Y$ such that each solution
of $\g$ is Lyapunov stable.
\end{prop}

\begin{proof}
This property obviously holds with respect to
the Riemannian fibre metric (\ref{ch80}) in Proposition
\ref{ch73} where $h=1$.
\end{proof}

Proposition \ref{ch75} can be improved as follows.

\begin{prop} \label{ch90} \mar{ch90}
Let $\la$ be a real number. Given a dynamic equation $\g$
defined by a complete vector field $\g$ (\ref{ch3}),
there is a Riemannian fibre metric on $Y$ such that
the Lyapunov spectrum of any solution of $\g$ reduces to  $\la$.
\end{prop}

\begin{proof}
Recall that the (upper) Lyapunov exponent of a
solution $s'$ with respect to
a solution $s$ is defined as the limit
\mar{ch51'}\beq
K(s,s')=\lim^-_{t\to\infty} \frac{1}{t}
\ln(\rho_t(s,s')). \label{ch51'}
\eeq
Let us provide $Y$ with the
Riemannian fibre metric (\ref{ch80}) in Proposition
\ref{ch73} where $h=\exp(\la t)$. A simple computation shows
that the Laypunov exponent (\ref{ch51'}) with
respect to this metric is exactly $\la$.
\end{proof}

If the upper limit
\be
\lim^-_{\rho_{t=0}(s,s')\to 0}K(s,s')=\la
\ee
is negative, the solution $s$ is said to be exponentially Lyapunov
stable. If there exists at least one positive Lyapunov exponent, one speaks
about chaos in a dynamical system \cite{gutz}.
Proposition \ref{ch90} shows that chaos in smooth dynamical systems
can be characterized in full by time-dependent Riemannian metrics.

\end{document}